\documentclass[journal=jacsat,manuscript=article]{achemso}

\usepackage{chemformula} 
\usepackage[T1]{fontenc} 

\usepackage{enumitem}

\author{Rair Mac\^edo}
\affiliation{SUPA School of Physics and Astronomy, University of Glasgow, Glasgow G12 8QQ, United Kingdom}
\email{Rair.Macedo@glasgow.ac.uk}
\author{Thomas Dumelow}
\affiliation[UERN]
{Departamento de F\'isica, Universidade do Estado do Rio Grande do Norte, Costa e Silva, 59625-620 Mossor\'o RN, Brazil}

\author{Robert E. Camley}
\affiliation[UCCS]
{Center for Magnetism and Magnetic Nanostructures, University ofColorado at Colorado Springs, Colorado Springs, Colorado 80918, USA}

\author{Robert L. Stamps}
\affiliation[UofM]
{Department of Physics and Astronomy, University of Manitoba, Winnipeg, MB R3T 2N2, Canada}

\title[]
  {Oriented Asymmetric Wave Propagation and Refraction Bending in Hyperbolic Media}

\abbreviations{IR,NMR,UV}
\keywords{American Chemical Society, \LaTeX}

\begin{document}







\begin{abstract}
Crystal quartz is a well-known anisotropic medium with optically active phonons in the THz region where hyperbolic phonon-polaritons can be excited. 
Here, we use this material to illustrate how the behavior of bulk and surface hyperbolic polaritons can be drastically modified by changing the orientation of the crystal’s anisotropy axis with respect to its surface. 
We demonstrate, both theoretically and experimentally, phenomena associated with the orientation of hyperbolic media. 
We show the consequences of slight changes in the crystal’s orientation in various ways, from the creation of hyperbolic surface phonon-polaritons to the demonstration of oriented asymmetric transmission of radiation passing through a hyperbolic medium. 
\end{abstract}

\section{Introduction}
$ $
Guided surface waves\cite{basov17,woessner15}, negative refraction,\cite{smith03a} and tunable focusing of THz radiation\cite{macedo16} are only a few of the intriguing optical effects recently enabled or enhanced by hyperbolic media, a class of materials in which the dielectric tensor or permeability tensor contains diagonal elements with different signs \cite{poddubny13}. 
The novel effects associated with this type of medium have triggered significant efforts devoted to the development of artificially engineered structures displaying hyperbolic dispersion\cite{hoffman07} in order to control and manipulate electromagnetic waves in unusual ways.

Widespread interest in this type of structure began with the upsurge of negative-index metamaterial research during the first decade of the millennium and the demand for easier and more efficient ways to achieve the same (or similar) novel optical phenomena as achieved with the early metamaterials \cite{smith03a,poddubny13}. 
Hyperbolic media were initially studied as an alternative route to achieving, for example, negative refraction; however, the area of polaritons in hyperbolic media has now taken its own direction \cite{poddubny13}. 
Natural materials are often at the center of such discussions, as in the case of several natural van der Waals crystals \cite{woessner15,dai18,dickson15} which are highly anisotropic compounds wherein atomic layers are coupled through van der Waals forces \cite{dai18}. 
These have recently gained in interest due to their unusual two-dimensional hyperbolic phonon-polariton propagation \cite{woessner15}. 
On the other hand, natural three-dimensional anisotropic crystals such as sapphire \cite{wang10} and calcite \cite{chen11} have also been studied as a route to hyperbolic dispersion with particularly impressive levels of transmission, suggesting that natural media are much more efficient than artificial structures.\cite{hoffman07,macedo14} 

In this work, we are interested in a less investigated aspect of hyperbolic media, the orientation of the anisotropy and how it can be used to engineer wave propagation. We focus on changes in the direction of the crystal's anisotropy with respect to the crystal's surface which, in general, will rotate the hyperbolic dispersion. This rotation modifies the behavior and frequencies where surface polaritons appear, changes the reflection coefficients and induces oriented symmetry breaking in the transmitted waves. In order to substantiate these claims, we calculate the infrared reflectivity and attenuated total reflection as well as transmission through a crystal quartz slab with different orientations for the anisotropy axes with respect to the surface, and show supporting experimental measurements. We also discuss the focusing by slab lenses, again as a function of orientation.

\section{Engineering Hyperbolic Dispersion through Anisotropy Orientation}

The symmetry of hyperbolic media is usually uniaxial i.e., two of the principal components of the dielectric tensor are equal \cite{dumelow16,rumpf15}. 
Here, we start by considering a crystal with its anisotropy axis along $z$ so that the permeability tensor is diagonal and has the form
\begin{equation}
\label{permtensor}
\stackrel{\leftrightarrow}{\varepsilon}(\omega) =
\begin{bmatrix}
\varepsilon_{\perp} & 0 & 0 \\
0 & \varepsilon_{\perp} & 0 \\
0 & 0 &  \varepsilon_{\parallel} \\
\end{bmatrix} 
=
\begin{bmatrix}
\varepsilon_{xx} & 0 & 0 \\
0 & \varepsilon_{yy} & 0 \\
0 & 0 &  \varepsilon_{zz} \\
\end{bmatrix}.
\end{equation}

The form of this tensor, however, will change if the anisotropy axis is rotated with respect to the surface by an angle $\varphi$, as shown in Figure~1(a) - details for $\stackrel{\leftrightarrow}{\varepsilon}(\omega)$ as a function of $\varphi$ are given in the Methods section. 
To illustrate the dramatic effects of orientation, we use crystal quartz, whose dielectric tensor components are shown in Figure~1(b), as the sample material. 
Even though this crystal has several phonon active modes spanning a broad range of the infrared spectrum (200-1200 cm$^{-1}$), many of which are potential candidates for hyperbolic dispersion, here we concentrate on only two of these modes, with resonances between 410 and 610 cm$^{-1}$, which are sufficient to illustrate our findings \cite{gervais75,silva12}.
\begin{figure}[h!]
\begin{center}
  \includegraphics[width=12cm]{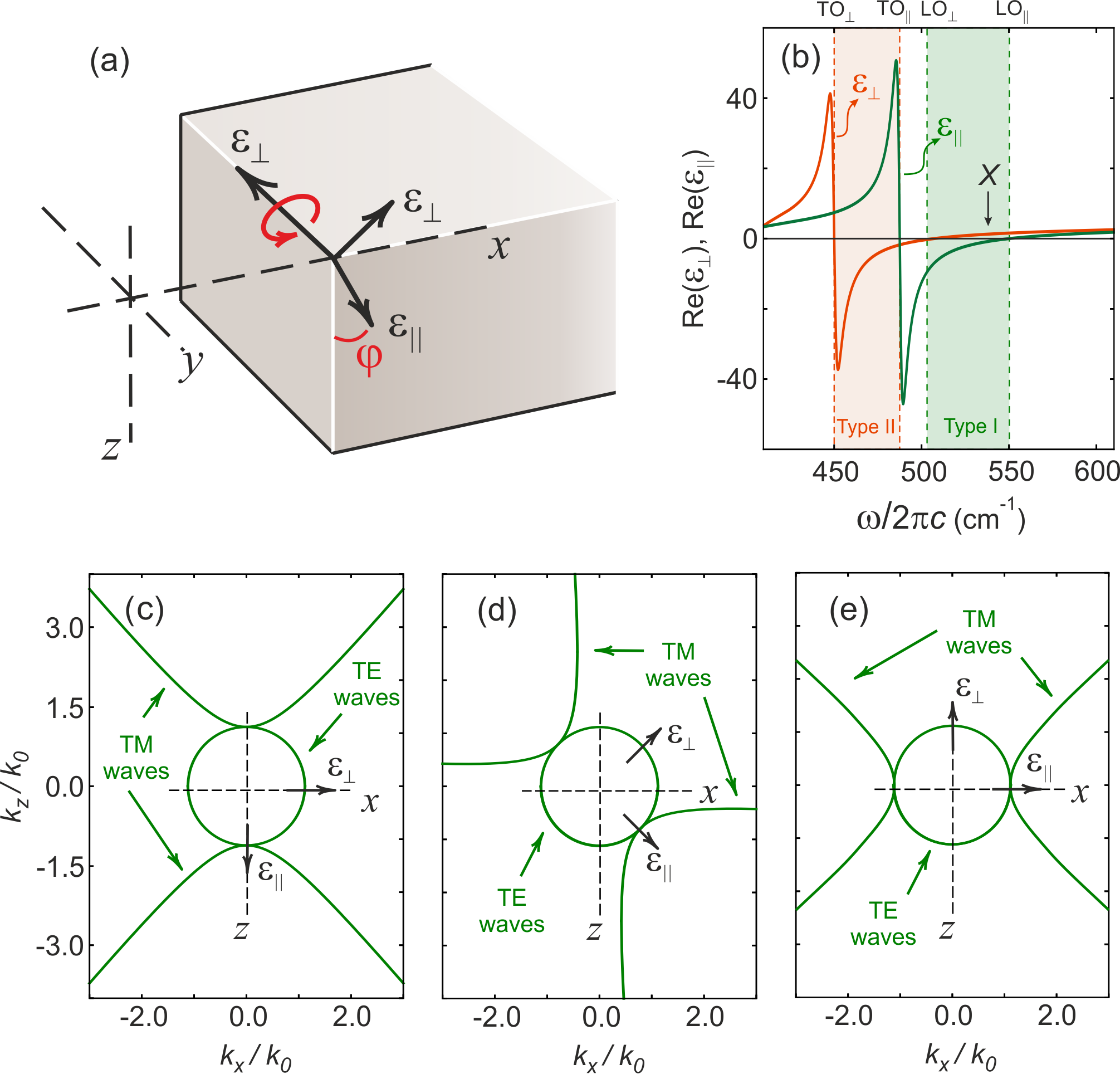}
  \caption{(a) Crystal geometry showing the direction of the anisotropy and plane of rotation used in this work. (b) Components of the dielectric tensor along the ordinary and extraordinary crystal's axes. Hyperbolic dispersion for TE and TM waves incident at frequency $X$ (537 cm$^{-1})$ on a quartz crystal's surface with anisotropy axis rotated by (c) $\varphi = 0$, (d) $\varphi = 45^\circ$ and (d) $\varphi = 90^\circ$. In part (b) LO and TO represent the frequencies of the longitudinal and transversal phonons respectively. }
\label{Geometry}
\end{center}
\end{figure}

Phonon polaritons can be excited by radiation incident on the surface from above the crystal.  
So far, however, our discussion has only described the behavior found in any simple anisotropic material, but such a crystal will only be hyperbolic if within some frequency range frequency its dielectric tensor components possess opposite signs. 
The regions in which $\varepsilon_{\perp}<0$ and $\varepsilon_{\parallel}>0$ are characterised as Type I hyperbolic regions and those with $\varepsilon_{\parallel}<0$ and $\varepsilon_{\perp}>0$ as Type II hyperbolic regions\cite{poddubny13,peragut17}, as indicated in Figure 1(b) by the green and orange shading respectively. 
The hyperbolic designation is a consequence of the shape of the medium's dispersion relation for excitation of polaritons by radiation incident in the $xz$ plane, with the electric field confined to the same plane (transverse magnetic (TM) polarization),
\begin{equation}
\frac{k_x^2}{\varepsilon_{\parallel}}+\frac{k_z^2}{\varepsilon_{\perp}} = k_0^2
\label{DispEq}
\end{equation}

where $k_x$ and $k_z$ represent the wavevector components in the plane of propagation and $k_0 = \omega/c$ is the amplitude of the free space wavevector at the excitation frequency $\omega$. 
The above dispersion relation yields symmetric open branch hyperbolas if the aforementioned conditions are met \cite{poddubny13} . 
In Figure 1(c) we show the TM hyperbolic dispersion at frequency $X$ ($\omega/2\pi c$ = 537 cm$^{-1}$), within the Type I region, where the characteristic hyperbolic shape is observed. 
For comparison the transverse electric (TE) curve, corresponding to the magnetic field in the $xz$ plane, leading to circular dispersion, is also shown.

If the optical axis is now rotated in the $xz$ plane by an angle $\varphi$, as defined in Figure~1(a), the hyperbolic dispersion curves also rotate, as shown in Figure~1(d) and 1(e). 
In the case of Type I hyperbolic dispersion with $\varphi=0$, propagation of TM polaritons is possible for all $k_x$ values, as seen in Figure~1(c). 
In addition, power flow is normal to the lower hyperbolic dispersion curve, leading to all-angle negative refraction \cite{silva12,maas16}. 
If $\varphi = 90^\circ$, where $\varepsilon_{\parallel}$  and $\varepsilon_{\perp}$ are effectively switched, only propagation of high $k_x$ waves will be allowed. 
The formalism of Equation (1) and (2), however, becomes more convoluted if the anisotropy axis no longer lined up with the Cartesian coordinate system, as in Figure~1(d). The dielectric tensor given by Equation~(1) is now modified, and the form of the tensor itself is no longer diagonal with $\varepsilon_{xz}=\varepsilon_{zx} \neq 0$ \cite{boardman15,urbas16}.
As noted earlier, the explicit changes in the dielectric tensor are given in the Methods section. 
The dispersion curves for the Type II hyperbolic case are very similar to those of the Type I dispersion, except that the curves for $\varphi = 0$ and $\varphi = 90^\circ$ are effectively swapped, so that only propagation for high $k_x$ is seen at $\varphi = 0$\cite{xu14} and all angle negative refraction is seen at $\varphi = 90^\circ$.  

The rotation of the optic axis, leading to a non-zero $\varphi$ has a significant impact on the behavior of waves propagating through the crystal as will be demonstrated in the following sections.  
Note that only the behavior of TM-polarized waves is affected by the rotation for this geometry. 

\section{Probing Bulk and Surface Hyperbolic Polaritons through Reflection }

We initially use simple reflection to investigate the effect of the anisotropy orientation. 
The reflection geometry is shown in Figure~2(a).  
TM polarized radiation is incident from a medium with higher dielectric constant than air (here we consider a diamond prism with dielectric constant $\varepsilon_p = 5.5$ above a semi-infinite quartz crystal) so that the radiation is allowed to couple with hyperbolic bulk polaritons i.e. propagating waves, with higher $k_x$, than possible when air is the incident medium. Here $k_x = \sqrt{\varepsilon_p}\sin\theta_1$ and $\theta_1$ is the incident angle. The reflectivity can then be calculated using the field continuity conditions at the interface.

\begin{figure}[h!]
\begin{center}
  \includegraphics[width=\linewidth]{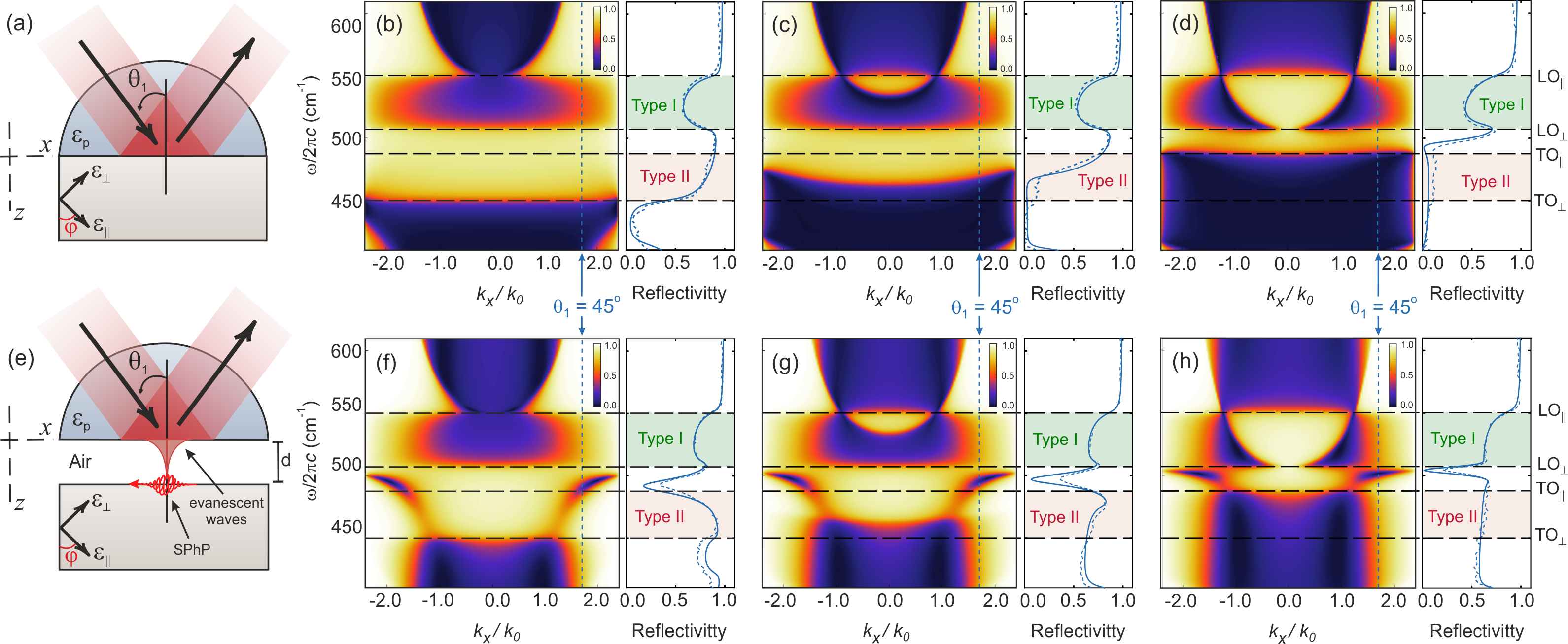}
  \caption{(a) Geometry for simple reflection where the radiation is incident from a diamond prism with dielectric constant $\varepsilon_p = 5.5$ onto a crystal quartz surface with its anisotropy axis rotated by an angle $\varphi$. 
  Reflection maps considering the geometry shown in (a) for rotation angles of (b) 0, (c) 45$^\circ$ and (d) 90$^\circ$. 
  The vertical dashed lines mark the reflection scan lines i.e. incident angle $\theta_1$ = 45$^\circ$. 
  The insets to the right of each reflection map show the equivalent experimental (dashed) and theoretical (solid) reflection lines at $\theta_1$ = 45$^\circ$. 
  (e) Geometry for attenuated total reflection (ATR) measurements where the prism and the crystal are separated by an air gap $d$. 
  Reflection maps for the ATR geometry considering rotation angles of (f) 0, (g) 45$^\circ$ and (h) 90$^\circ$ and their respective ATR scan lines at $\theta_1$ = 45$^\circ$ so that the radiation couples with surface phonon polaritons (SPhP's). Here the air gap is taken as $d$ = 1.5~$\mu$m.}
\label{Ref_ATR_Maps}
\end{center}
\end{figure}

In Figure~2(b)-(d) we show reflection maps for various rotation angles of the crystal's anisotropy. 
These are calculated reflectivity plots as a function of both frequency and $k_x$ for the anisotropy axis at various angles with respect to the crystal surface. 
This is a particularly useful technique as it can be used to trace the boundaries between bulk bands (where the reflectivity is low) and nonpropagating regions (where the reflectivity is large). 
In Figure~2(b) we show the reflection map at $\varphi = 0$, so that the crystal's extraordinary axis lies along $z$. 
In the Type I hyperbolic regions, one sees a large $k_x$ range where bulk polaritons can propagate. 
In contrast, in the Type II regions there is no propagation within the $k_x$ range shown. 
In these areas, large reflections can be seen. 
This behavior is confirmed by the excellent agreement between the theoretical and experimental reflection spectra shown in the inset to the right of Figure~2(b) for an incident angle of $+45^\circ$.

The regions in which we find either propagating or reflected waves are extremely sensitive to changes in the anisotropy axis direction, as seen by comparing the $\varphi =0$ results (Figure~2(b)) with those for $\varphi = 45^\circ$ (Figure~2(c)). 
In the latter case, a small nonpropagating region, with high reflectivity, is created in the Type I hyperbolic region centered around $k_x = 0$. 
This nonpropagating region expands as $\varphi$ increases, reaching a maximum width at $\varphi = 90^\circ$ as shown in Figure~2(d). 
In some sense this is similar to the gaps in propagation introduced by a periodic structure, except that here the gap is introduced through a different mechanism which does not require a structured material \cite{lopez06}. 
The theoretical and experimental reflectivity curves again show good agreement for an incident angle of $+45^\circ$, i.e. along the dotted line in the reflection maps. 
It is also important to note that the non-propagating region associated with Type II hyperbolic dispersion at lower frequencies gets narrower as the rotation increases. In this region the reflectivity for an incident angle of $+45^\circ$ is significantly affected.

The regions in which propagation into the crystal is forbidden are of particular interest as they are where surface phonon polaritons (SPhP's) can exist. 
Here, we employ the attenuated total reflection (ATR) technique as a way to investigate the existence and behavior of these SPhP's. 
ATR been proven to be an excellent tool to probe the behavior of surface polaritons in anisotropic dielectric\cite{lee15} as well as magnetic media\cite{abraha96}. 
We use an ATR setup in the classic Otto configuration shown in Figure~2(e). 
Here a prism is placed above the crystal's surface, separated by an air gap of distance $d$. 
By adjusting the distance between the prism and the crystal surface it is possible to make the incident electromagnetic waves couple to surface polaritons. 

We show the ATR reflection maps in Figure~2(f)-(h). 
In the ATR configuration, evanescent waves within the air gap couple with surface polaritons, which can be seen in the regions where no bulk wave propagation is allowed.
They usually appear as sharp dips in reflectivity, and such dips are clearly seen in spectra shown in Figure~2(f), (g) and (h). 
Note how the frequencies at which the surface phonon-polaritons appear shift with the rotation of anisotropy axis and at $\varphi=90^\circ$ the surface phonon-polariton ends up confined to a very narrow region, between the Type I and Type II hyperbolic regions, where both $\varepsilon_{xx}$ and $\varepsilon_{zz} <0$.

It is worth noting that both the reflectivity and ATR reflection maps are symmetrical around $k_x = 0$, showing that reflectivity is independent of the sign of the incident angle $\theta_1$. 
This is in accordance with the Helmoltz reciprocity principle by which, in the absence of a magnetic field, the position of the source and detector in a particular optical setup can be exchanged without altering the observed intensity.

\section{Oriented Asymmetric Wave Propagation}

Up to this point, we have only discussed the effect of arbitrary orientation of the anisotropy axis in terms of the behavior of reflected waves from a semi-infinite surface in the spectral region of interest. 
However, transmitted waves can also be highly affected by the direction of the anisotropy. 
To illustrate this, we will now look at the behavior of infrared radiation travelling through a flat crystal quartz slab, surrounded by air. 

In Figure~3(a) we show the calculated transmission map, for the same frequency range as used for investigating reflection behavior, for incident angles between $-90^\circ$and $+90^\circ$ (within the light line of air), for a crystal whose anisotropy axis is perpendicular to the surface ($\varphi = 0$). 
Large transmission can be seen in the frequency region where Type I hyperbolic phonon-polaritons are found, as may be expected from Figure 1(c). 
In fact, this transmission is far higher than observed for equivalent artificial structures \cite{hoffman07}, in agreement with previous works on natural hyperbolic crystals \cite{macedo14,dumelow16}.

\begin{figure}[h!]
\begin{center}
  \includegraphics[width=0.8\linewidth]{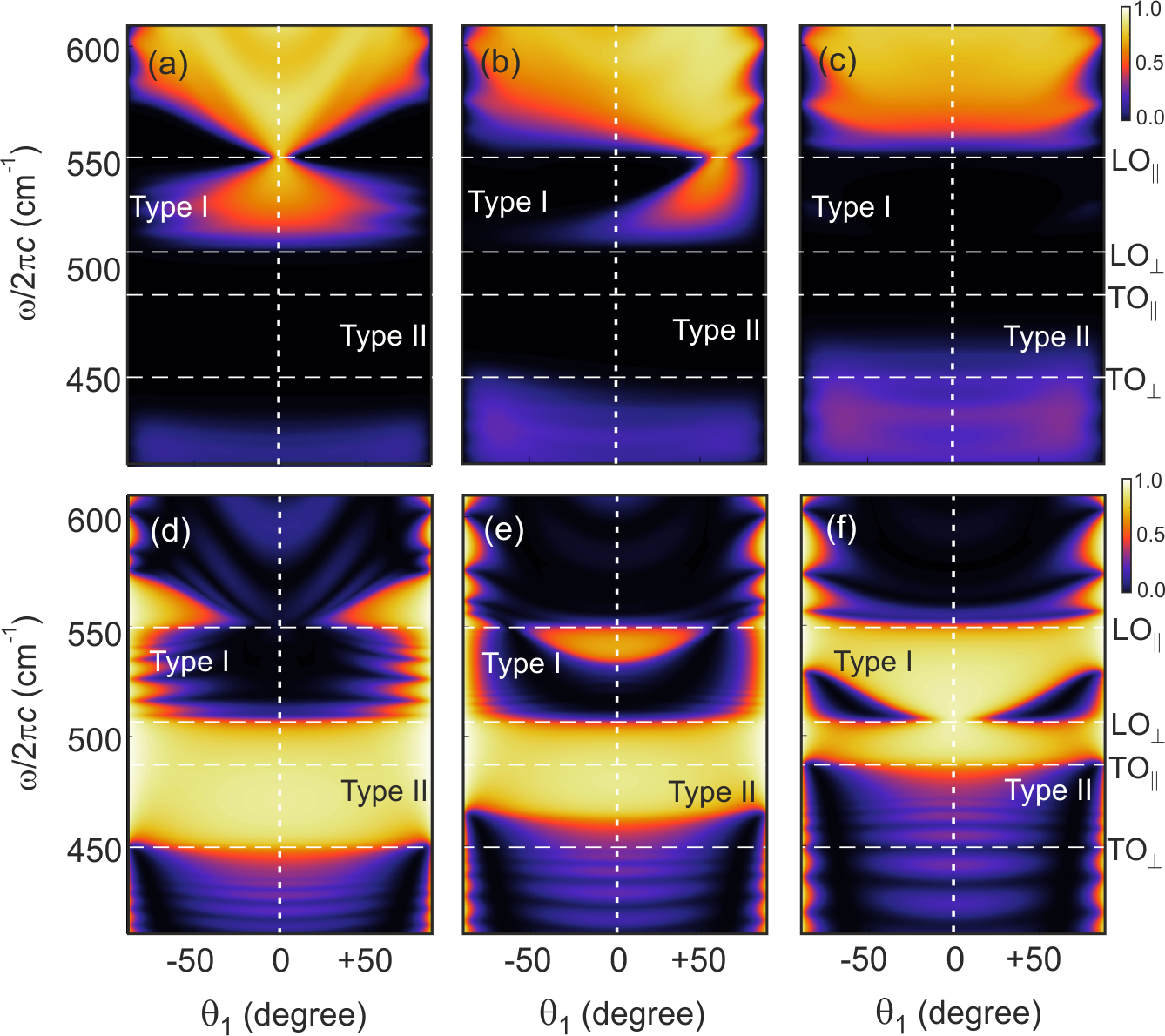}
  \caption{(a)-(c) Transmissivity and (d)-(f) reflectivity for electromagnetic waves incident onto a 25~$\mu$m thick quartz crystal slab as a function of both frequency and incident angle. 
  The anisotropy axis directions are (a) and (d)$\varphi=0$, (b) and (e) $\varphi = 45^\circ$, and (c) and (f)$\varphi = 90^\circ$ - as defined in Fig 1(a).}
\label{Ref_Trans_Maps}
\end{center}
\end{figure}

In Figure~3(b) and 3(c), we show the effect of rotating the crystal axes with respect to the surface. 
The most intriguing behavior, perhaps, is when the optical axis is rotated by $45^\circ$, as shown in Figure 3(b). 
In stark contrast to the case when $\varphi = 0$, here the transmission of waves is extremely asymmetric i.e. the transmission level for a given incident angle differs from that of its negative counterpart. 
For example, for the high rotation angle shown ($\varphi = +45^\circ$), considering propagation to be downwards, there is hardly any transmission in the Type I region for negative incident angles, but fairly high transmission for high positive angles. 
This symmetry breaking depends on the direction of the rotation, however, so that for $\varphi = -45^\circ$, the transmission would yield the mirror image plot of what is shown in Figure~3(b). 
Note that when the anisotropy axis is rotated by $90^\circ$ the transmission is again symmetric (see Figure~3(c)).
However, the behavior is opposite to that seen in Figure~3(a), i.e. the hyperbolic Type I and Type II phonon-polariton behaviors seem to be now switched so that there is some propagation in within the Type II region but none in the Type I region.

While the transmission through a thin slab is highly affected by the orientation of the anisotropy, showing distinct asymmetry for $\varphi = +45^\circ$, the reflectivity remains symmetric for all $\varphi$ as shown in Figure~3(d)-(f). 
These reflectivity maps show very similar features to those shown in Figure 2~(b)-(d) for the range $-1\leq k_x/k_0\leq +1$, corresponding to an incident angle range of $-90^\circ$ to $+90^\circ$ in air. 
In terms of the Helmholtz reciprocity, we can see that in reflection reversing the sign of the incident angle is equivalent to exchanging the positions of the source and detector, but in transmission this is not in general the case, so asymmetrical transmission is allowed. 
Physically, we can interpret such asymmetry as a consequence of the transmitted waves being affected by damping in different ways according to the sign of the incident angle. 
For instance, in the present case, the beam path of negatively incident beams within the Type I region is longer than that of its positive counterpart, which gives rise to the oriented asymmetry for the transmitted beam but does not affect the reflected waves. 

This oriented asymmetry of the transmitted waves ought to have a significant impact on optical effects such as negative refraction, slab lensing, focusing and imaging recently studied in hyperbolic media. 
In order to investigate the implications of this rotation we first look at the behavior of waves travelling through a crystal quartz slab for the unrotated crystal ($\varphi=0$) geometry. 
In Figure~4(a) the angle of refraction is given as a function of both incident angle (from $-50^\circ$ to $+50^\circ$) and frequency. 
This angle, which represents the direction of power flow within the crystal, is, in the absence of damping, perpendicular to the relevant dispersion curve, as mentioned in the introduction. 
A more general method for obtaining the direction is to calculate the Poynting vector ($\mathbf{S=E\times H^*}$). 
The angle of refraction can be obtained from the components of the time averaged Poynting vector$\langle \mathbf{S} \rangle = \frac{1}{2}$Re($\mathbf{S}$) in the incidence plane and is given by $\tan\theta_2 = {\langle S_{2x}\rangle}/{\langle S_{2z}\rangle}$ \cite{macedo16,dumelow93}. 
In the Type II hyperbolic region there is no propagation into the crystal. 
Instead, the power flow is along the surface, corresponding to $\theta_2=\pm 90^\circ$, with exponential decay into the crystal (although there are some small deviations from this behavior in the presence of damping). As discussed earlier, only high-$k_x$ waves, outside the range shown in the Figure, are allowed to propagate into the crystal in these regions. 
In the range shown, evanescent decay into the crystal dominates, any power flow into the crystal being associated with absorption losses due to damping. 
In the Type I hyperbolic region, not only is propagation into the crystal possible for all incident angles (as shown in Figure~4(a)), but so is all-angle negative refraction. 

\begin{figure}[h!]
\begin{center}
  \includegraphics[width=0.8\linewidth]{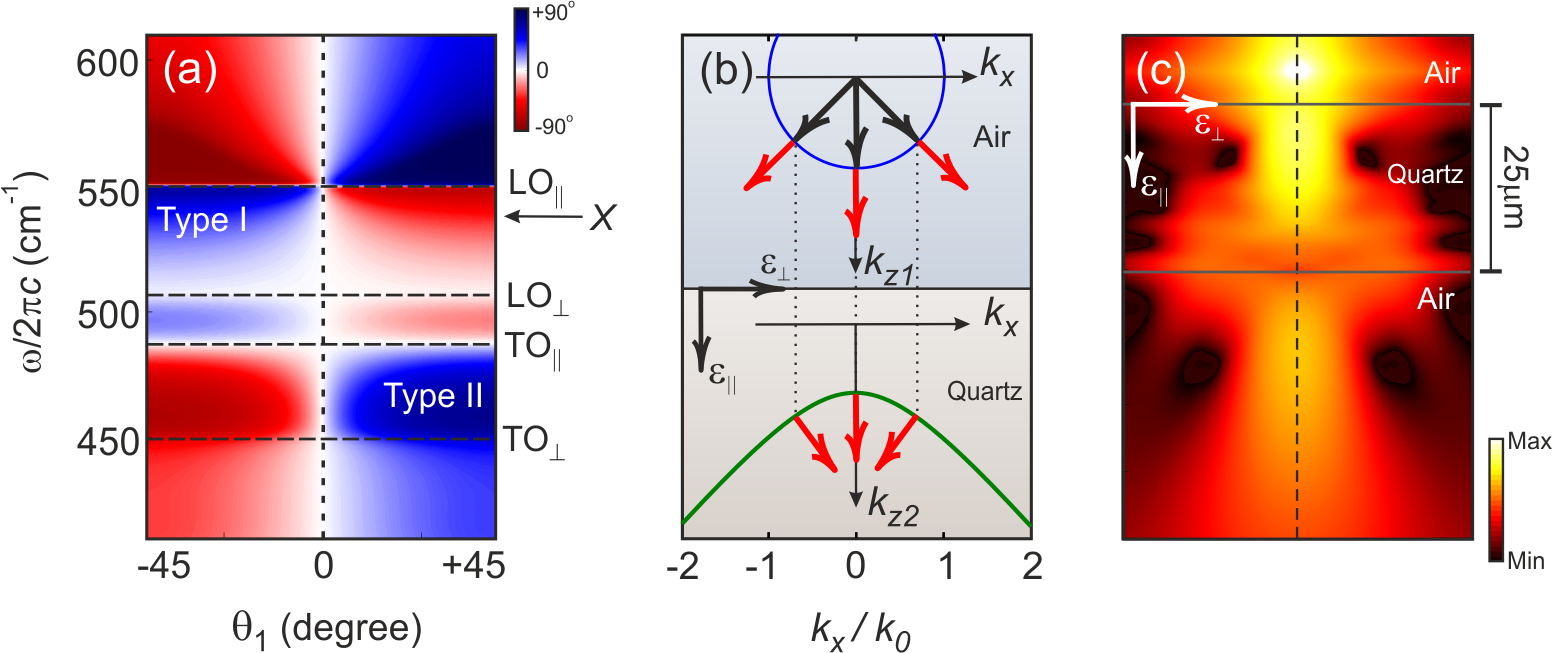}
  \caption{(a) Calculated angle of refraction as function of both frequency and incident angle (from $-50^\circ$ to $+50^\circ$) for crystal quartz with $\varphi = 0$, so that $\varepsilon_\parallel$ lies along $z$. 
  (b) Schematics of optical rays incident onto a crystal quartz surface and their direction of propagation due to the crystal's hyperbolic dispersion. 
  (c) Power flow intensity showing image formation due to a line source irradiating TM-polarized radiation in all directions placed above a 25~$\mu$m quartz crystal slab.
  The frequency of the incident radiation for (b) and (c) is $X$ (537 cm$^{-1}$: see Figure 1(b)) which falls within the Type I hyperbolic region.}
\label{Slabs0}
\end{center}
\end{figure}

In Figure~4(b) we show a schematic of how such refraction occurs, showing beam propagation perpendicular to the relevant dispersion curve.  
If the incident rays all stem from a line source along $y$, whose lateral position we take as $x = 0$, above the slab, then each positive incident ray will meet its negative counterpart within the slab at $x = 0$. In Figure~4(c) we show focusing by a quartz flat slab under the same conditions as given in Figure~4(b). 
Here all beams are focused both inside and outside the slab in a similar manner to that which occurs in Veselago lenses\cite{veselago68} using a negative index slab.

The efficiency of these devices is highly connected to efficient transmission and, since high angles of rotation such as $45^\circ$ reduce the transmittance considerably, we concentrate the following discussion on the behavior of transmitted waves through a crystal quartz slab whose anisotropy axis rotated by only $5^\circ$ (details on sample preparation and control of the anisotropy axis direction are given in the Methods Sec. 5.3). 
We start by looking at the measured transmissivity as function of incident angle and frequency, as shown in Figure~5(a). 
This may seem qualitatively similar to that shown in Figure~4(a) when the anisotropy is parallel to the $z$ axis. 
However, the transmissivity is significantly asymmetric despite such a small rotation. 
The plot is in excellent agreement with Figure~4(b) where we show the calculated transmissivity. 
Note that the greatest asymmetry is observed in the Type I hyperbolic region.    

\begin{figure}[h!]
\begin{center}
  \includegraphics[width=\linewidth]{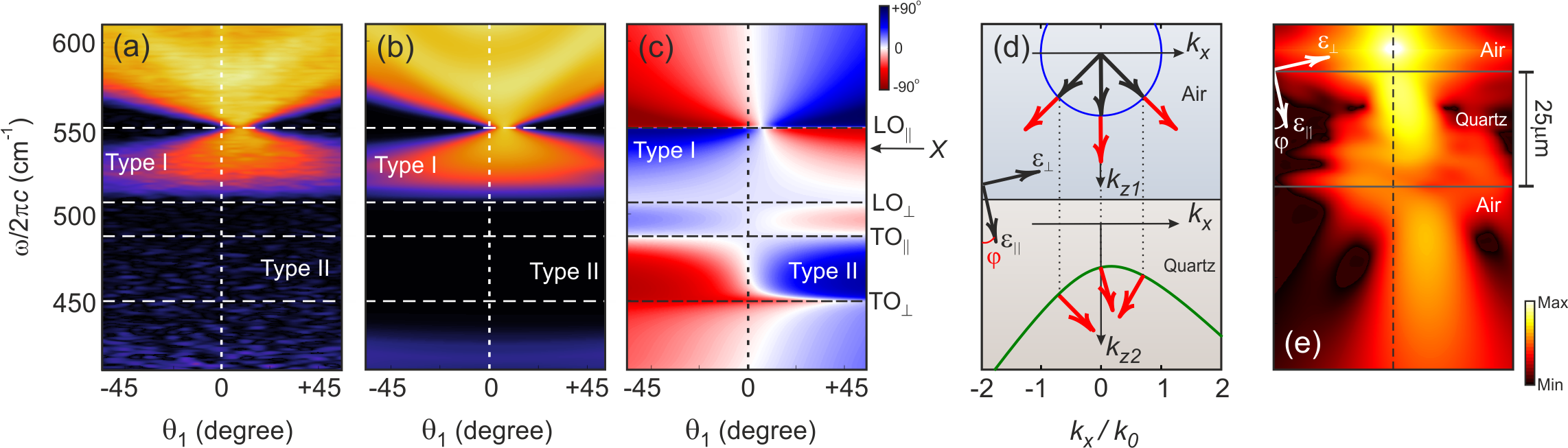}
  \caption{(a) Experimental and (b) calculated transmissivity as function of both frequency and incidence angle (from $-50^\circ$ to $+50^\circ$) for a 25~$\mu$m crystal quartz slab when $\varphi = 5^\circ$. 
  (c) Calculated angle of refraction under the same conditions as for part (a). 
  (d) Schematics of optical rays incident onto a crystal quartz surface and their direction of propagation according to the crystal's hyperbolic dispersion for $\varphi = 5^\circ$. 
  (e) Calculated TM-polarized image formation due to a line source placed above a 25~$\mu$m crystal quartz slab for $\varphi = 5^\circ$. 
  The frequency of the incident radiation for (d) and (e) is $X$ (537 cm$^{-1}$) which falls within the Type I hyperbolic region.}
\label{Slabs5}
\end{center}
\end{figure}

In Figure 5(c) we plot the angle of refraction for $\varphi = 5^\circ$. Despite such a small rotation, the results for this angle are completely different from those shown in Figure~4(a). 
Note that the most significant asymmetry is found in regions where hyperbolic behavior exists. 
However, some asymmetry is also seen in the other regions.
The effect of this asymmetry on both the transmission levels and angle of refraction can be understood when looking at the behavior of propagating beams. 
In Figure~5(d), we show the refraction of rays according to the hyperbolic dispersion (which is now slightly rotated). 
This gives rise to the refraction bending i.e, positive and negative incident angles (here we show $\pm 45^\circ$) no longer have equivalent angles of refraction. 
This leads to a displacement of the focal point as the rays do not meet at $x = 0$ within the slab any more. 
In Figure~5(e) we show power flow behavior, leading to image formation, due to a line source placed above a crystal quartz slab for $\varphi = 5^\circ$. 
The beams are once more focused inside and outside the slab, but with a significant lateral displacement in a direction which depends on the sign of $\varphi$. 
It is also important to note that, due to the rotation of all propagating beams, the left-hand region (for the positive $\varphi$ case shown in the figure) below the slab is now completely dark, and no beam is allowed to travel in this region.

\subsection{Final Remarks}

The examples shown above display a few of the possibilities offered by hyperbolic polaritons. 
While we have concentrated on a particular rotation geometry, many more possibilities should arise using other anisotropy orientations. 
For instance, a geometry where the anisotropy is allowed to rotate out of the plane as well as in plane offers interesting possibilities for changing the frequencies of surface hyperbolic polaritons. 
In this case, the off-diagonal elements induced by a non-zero $\varphi$ do not lie in the same plane as the incident electric fields of the radiation, and the hyperbolic behavior and its features will be affected accordingly.

As seen from the angle of refraction plots, the regions where propagation along the surface is observed are drastically affected by the anisotropy orientation, and in the frequency range of Type I hyperbolic behavior their range increases with the rotation angle $\varphi$. 
In these regions, optical effects such as Goos-H\"anchen shifts\cite{macedo13,wu18} should occur and such effects should depend on the direction of the anisotropy.

While we have shown how hyperbolic surface-phonon polaritons behave as the anisotropy direction is rotated in natural dielectric crystal quartz, we expect that our findings can be applied to any hyperbolic system. 
For instance, in magnetic crystals the behavior of surface magnon-polaritons ought to be even more captivating as the rotation of the magnetic easy axis combined with an external field can induce nonreciprocity of both bulk and surface modes \cite{camley82,jensen97}. 
Sampling areas of non-propagation in magnetic crystals, similar to those shown in Figure~2(b)-(d), is also of particular interest. 
For example, in magnetic ferrites\cite{harward14} and magnetic multilayers \cite{fal11,macedo18b}, non-propagating regions have been used to engineer band stop filters at GHz frequencies to be used in signal processing.  
Even though the form of the dielectric tensor here has some similarities with the permeability tensor of magnetic crystals in presence of an external field, the nature of its off-diagonal components completely differs. 
In the present case, the off-diagonal components are equal to one another ($\varepsilon_{xz}=\varepsilon_{zx}$) and, in the absence of damping, are wholly real, which leads to the effects discussed in this letter. 
In magnets, however, the nature of the off-diagonal components is distinctively different, in which case additional nonreciprocal effects, associated with the particular form of the permeability tensor, may further modify the behavior \cite{macedo17,camley87,MacedoCamley18}.
We believe that our findings on the oriented asymmetric transmission can also be used as a potential avenue for applications on optical devices such as direction dependent electromagnetic filters.   

In this work, we have concentrated on crystal quartz as the sample material, where the change in the anisotropy orientation is mediated through different crystal cuts.
However, the orientation of the optical axis can be easily tuned in media such as liquid crystals where a static electric field may be externally applied to orient the crystal internal structure along a certain direction. 
A similar approach has been recently suggested for switching between negative and positive refraction by changing the dispersion shape from hyperbolic to elliptic \cite{cao16,pawlik14}.
We believe that an electric field can be used to choose whether propagation is allowed into the liquid crystal structure or in its surface\cite{spinozzi11} and therefore control not only the nature of propagation but tune lateral shifts on reflection or the focal point of a flat lens. 
Moreover, our findings can be also used to understand and optimize cloaking devices\cite{pawlik12} where the invisibility is mediated by changes in the structure's local fields, and hence the shape of the dispersion curves.  

Finally, with the recent interest in hyperbolic polaritons in two-dimension van der Waals structures \cite{gilburd17,li15,caldwell14}, the findings presented here for three-dimensional natural crystals can be directly translated into such two-dimensional structures if the direction of the anisotropy in those structures is controlled. 
Our findings can potentially help the performance of guiding waves as well as to control the direction of near-field distribution of hyperbolic polaritons in meta-surfaces launched by antennas \cite{yang17,dai15}.   

\section{Methods}
\subsection{Axis Transformation}

The rotation of the anisotropy direction discussed throughout this work and illustrated in Figure~1(a) can be represented through a crystal axis transformation \cite{lee14,al15}.
We introduce a rotation in the $xz$ plane given by an angle $\varphi$.
The dielectric tensor components have to be modified in the new coordinate system and this can be done by $\stackrel{\leftrightarrow}{\varepsilon'}(\omega) = \mathbf{T}\stackrel{\leftrightarrow}{\varepsilon}(\omega)\mathbf{T}^{-1}$, where $\stackrel{\leftrightarrow}{\varepsilon}(\omega)$ is the original dielectric tensor, $\mathbf{T}$ is a transformation matrix in the $xz$ plane and $\mathbf{T}^{-1}$ is its transpose \cite{al15}. 
The new dielectric tensor components are given by

\begin{subequations}
\begin{align}
\varepsilon_{xx}=\varepsilon_{\perp}\cos^2\varphi+\varepsilon_{\parallel}\sin^2\varphi \\
\varepsilon_{zz}=\varepsilon_{\perp}\sin^2\varphi+\varepsilon_{\parallel}\cos^2\varphi \\
\varepsilon_{xz} = \varepsilon_{zx} = (\varepsilon_{\parallel}-\varepsilon_{\perp}) \cos\varphi\sin\varphi\\
\end{align}
\label{epsPrime}
\end{subequations}  
Each original component of the dielectric tensor can be modeled as\cite{silva12}

\begin{equation}
\varepsilon_u = \varepsilon_{\infty,u}\prod_i\left(
\frac{\omega^2_{Ln,u}-\omega^2-i\omega\gamma_{Ln,u}}{\omega^2_{Tn,u}-\omega^2-i\omega\gamma_{Tn,u}}
\right),
\label{epsComps}
\end{equation}
where $u$ represents either $\parallel$ or $\perp$. Here $\omega$ is the frequency, $\varepsilon_{\infty,u}$ is the high frequency dielectric constant, $\omega_{Tn,u}$ is the frequency of the transverse optical (TO) phonons, $\omega_{Ln,u}$, is the frequency of the longitudinal optical (LO) phonons and $\gamma_{Ln,u}$ and $\gamma_{Tn,u}$ are the damping parameter of the $n-$th LO or TO phonon mode respectively.  
The specific parameters we use in Eq (4) are given in Ref \cite{silva12}

We can now apply Maxwell's equations with the transformed dielectric tensor given in Equation (3) to find the dispersion relation for TM radiation within the crystal to be

\begin{equation}
\varepsilon_{zz}k_z+\varepsilon_{xx}k_x+2\varepsilon_{xz}k_zk_x=\frac{\omega^2}{c^2}(\varepsilon_{xx}\varepsilon_{zz}-\varepsilon_{xz})
\label{DispPhi}
\end{equation}
 
\subsection{Transmission and Reflection Through Layers}

In order to calculate the transmission and attenuated total reflection of waves on the surface of crystal quartz we employ a standard transfer matrix technique. 
This method has been detailed elsewhere\cite{dumelow93b,macedo17} so here we only summarize the main aspects of it. 
This technique consists of matching the boundary condition for the fields propagating up and down at each boundary. 
For dielectric crystals, as it is our case, it is convenient to work with the magnetic field propagating through the layers. 
This field can be written as a sum of waves propagating upwards and downwards as follows
\begin{equation}
\mathbf{H_y} = a_{nl}e^{-ik_{nz}z}+b_{nl}e^{ik_{nz}z}
\label{HyDown}
\end{equation}
or
\begin{equation}
\mathbf{H_y} = a_{nu}e^{-ik_{nz}z}+b_{nu}e^{ik_{nz}z}
\label{HyUp}
\end{equation}
where $z$ is the position with respect to either the bottom ($l$) or top ($u$) of the $n-$th layer. These can be related as 
\begin{equation}
\label{abMatrix}
\begin{bmatrix}
a_{nu} \\
b_{nu} \\
\end{bmatrix}
=\mathbf{F}
\begin{bmatrix}
a_{nl} \\
b_{nl} \\
\end{bmatrix}.
\end{equation}
where
\begin{equation}
\label{FMatrix}
\mathbf{F} = 
\begin{bmatrix}
e^{-ik_{nz}d} & 0\\
0 & e^{ik_{nz}d}\\
\end{bmatrix}.
\end{equation}
for a layer of thickness $d$. For a three-layered system, the relation between the fields in the first and third layer, the first layer being the incident layer, can be written as
\begin{equation}
\label{refMatrixFinal}
\begin{bmatrix}
a_{1u} \\
b_{1u} \\
\end{bmatrix}
=\mathbf{M_1}\times\mathbf{F}\times\mathbf{M_2}
\begin{bmatrix}
a_{3l} \\
0 \\
\end{bmatrix},
\end{equation}
where $\mathbf{M_1}$ and $\mathbf{M_2}$ are the matrices relating the fields on either side of the first and the second interface respectively. For instance, for the ATR geometry the first interface is Prism/Air and the second Air/Quartz, and for transmission we have Air/Quartz first and Quartz/Air second. The above matrix equation can be summarized as:
\begin{equation}
\label{refMatrix}
\begin{bmatrix}
a_{1u} \\
b_{1u} \\
\end{bmatrix}
=
\begin{bmatrix}
R_{11} & R_{12}\\
R_{21} & R_{22}\\
\end{bmatrix}
\begin{bmatrix}
a_{3l} \\
0 \\
\end{bmatrix}
\end{equation}
so that the complex reflection coefficient is given by
\begin{equation}
r = \frac{R_{21}}{R_{11}}
\end{equation}
and the complex transmission coefficient is
\begin{equation}
t = \frac{1}{R_{12}}.
\end{equation}
Note that throughout this work we have used the transmissivity and reflectivity intensities which are given by $tt^*$ and $rr^*$ respectively.

\subsection{Experimental Set Up and Parameters}

The parameters for the infrared phonon active modes and their absorptions used through this work to model the wave propagation in crystals quartz were initially extracted from F. Gervais and B. Piriou’s work.\cite{gervais75} 
These have been slightly modified by Estevam et al,\cite{silva12} whose parameters were ultimately used in our simulations for a somewhat better fit to the experimental data. 
These parameters were then plugged into Eq.~(4) to obtain the permeability tensor components as a function of frequency.

All experiments were performed using a Bruker Vertex 70 far-infrared spectrometer with a resolution of 2~cm$^{-1}$ and each spectrum was averaged 16 times. 
For the reflectivity and attenuated total reflection measurements, we used a Bruker diamond ATR unit with a diamond prism and incident radiation at an angle of $45^\circ$. We used a KRS-5 infrared polarizer placed in the beam path in order to obtain TM-polarized waves for all measurements. 
The samples used were crystal quartz obtained from Boston Piezo Optics Inc. 
These were chemically polished where a particular orientation of the anisotropy axis was chosen with respect to the surface. 
This anisotropy orientation was precisely determined by single-crystal X-ray diffraction measurements where a crystal is mounted on a goniometer which allows the positioning of the crystal at selected orientations. 
For the reflection measurements we used 10~mm thick slabs of 20~mm diameter. 
For transmission we used samples with the same diameter but 25~$\mu$m thick.

\begin{acknowledgement}

This work was financially supported by the Leverhulme Trust, SUPA, and the University of Glasgow through LKAS funds. 
R. Macedo would also like to acknowledge useful discussions with A. D. Boardman and the hospitality of The University of Colorado at Colorado Springs where the experimental work was performed - D. Bueno-Baques and O. Melnik, in particular, for assistance with the experimental set up. 
T. Dumelow acknowledges support of the Brazilian agency CNPq and R. L. Stamps acknowledges the support of the Natural Sciences and Engineering Research Council of Canada (NSERC) - R. L. Stamps a été financée par le Conseil de recherches en sciences naturelles et en génie du Canada (CRSNG). 

\end{acknowledgement}


%


\bibliography{Ref_QuartzACS}

\section*{Graphic Table of Content}
\begin{figure}[h!]
\begin{center}
  \includegraphics[width=12cm]{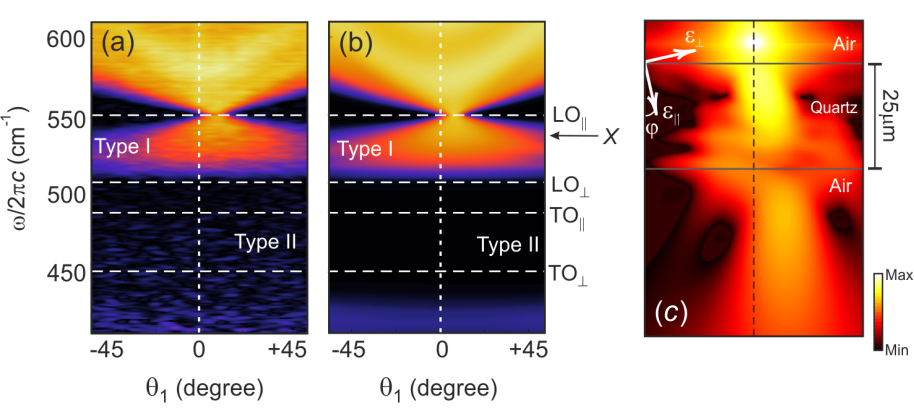}
\end{center}
\end{figure} 
\end{document}